\documentclass[prb,showpacs,preprintnumbers,amsmath,superscriptaddress,twocolumn]{revtex4} %
\usepackage{mathrsfs}%,amssymb
\usepackage{graphicx}% Include figure files
\usepackage{dcolumn}% Align table columns on decimal point
\usepackage{bm}% bold math
\usepackage{amsmath}
\usepackage{color}
\makeatletter
\renewcommand{\section}{\@startsection{section}{1}{0mm}
  {-\baselineskip}{0.5\baselineskip}{\bf\leftline}}
\makeatother

\makeatletter
\renewcommand{\subsection}{\@startsection{subsection}{1}{0mm}
  {-\baselineskip}{0.5\baselineskip}{\bf\leftline}}
\makeatother

\makeatletter
\renewcommand{\subsubsection}{\@startsection{subsubsection}{1}{0mm}
  {-\baselineskip}{0.5\baselineskip}{\bf\;\leftline}}%
\makeatother

\begin{document}

\preprint{APS/123-QED}
\title{First principles study and empirical parametrization of twisted bilayer $\textrm{MoS}_2$ based on band-unfolding}% Force line breaks with \\

\author{Yaohua~Tan}
\email{tyhua02@gmail.com}
%\affiliation{%
%School of Electrical and Computer Engineering,Network for
%Computational Nanotechnology,Purdue University, West Lafayette,
%Indiana 47906, USA % \textbackslash\textbackslash
%}
\affiliation{%
Department of Electrical and Computer Engineering, University of Virginia, Charlottesville,
Virginia 22904, USA% \textbackslash\textbackslash
}

\author{Fan~Chen}
\affiliation{%
School of Electrical and Computer Engineering,Network for
Computational Nanotechnology,Purdue University, West Lafayette,
Indiana 47906, USA
}
%\author{Gerhard~Klimeck }
%\affiliation{%
%School of Electrical and Computer Engineering,Network for
%Computational Nanotechnology,Purdue University, West Lafayette,
%Indiana 47906, USA  % \textbackslash\textbackslash
%}

\author{Avik~W. Ghosh}
\affiliation{%
Department of Electrical and Computer Engineering, University of Virginia, Charlottesville,
Virginia 22904, USA% \textbackslash\textbackslash
}
%\author{Michael~Povolotskyi}
%\affiliation{%
%School of Electrical and Computer Engineering,Network for
%Computational Nanotechnology,Purdue University, West Lafayette,
%Indiana 47906, USA % \textbackslash\textbackslash
%}
%\author{Tillmann~Kubis }
%\affiliation{%
%School of Electrical and Computer Engineering,Network for
%Computational Nanotechnology,Purdue University, West Lafayette,
%Indiana 47906, USA % \textbackslash\textbackslash
%}
%\author{Timothy B.~Boykin} \affiliation{
%University of Alabama in Huntsville,Huntsville, Alabama 35899, USA % with \\
%}%
%\author{Gerhard~Klimeck }
%\affiliation{%
%School of Electrical and Computer Engineering,Network for
%Computational Nanotechnology,Purdue University, West Lafayette,
%Indiana 47906, USA  % \textbackslash\textbackslash
%}
%
%%\altaffiliation[Also at ]{Physics Department, XYZ University.}
%
%
%%% \homepage{http://www.Second.institution.edu/~Charlie.Author}

%
%\author{Ann  Author}
% \altaffiliation[Also at ]{Physics Department, XYZ University.}%Lines break automatically or can be forced with \\
%\author{Second Author}%
% \email{Second.Author@institution.edu}
%\affiliation{%
%Authors' institution and/or address\\
%This line break forced with \textbackslash\textbackslash
%}%
%
%\author{Charlie Author}
% \homepage{http://www.Second.institution.edu/~Charlie.Author}
%\affiliation{
%Second institution and/or address\\
%This line break forced% with \\
%}%

\date{\today}% It is always \today, today,
             %  but any date may be explicitly specified
\begin{abstract}
We explore the band structure and ballistic electron transport in twisted bilayer $\textrm{MoS}_2$ using Density Functional Theory (DFT). The sphagetti like bands are unfolded to generate band structures in the primitive unit cell of the original un-twisted $\textrm{MoS}_2$ bilayer and projected onto an individual layer. The corresponding twist angle dependent indirect bandedges are extracted from the unfolded band structures. Based on a comparison within the same primitive unit cell, an efficient two band effective mass model for indirect conduction and valence valleys is created and parameterized by fitting the unfolded band structures. With the two band effective mass model, transport properties - specifically, we calculate the ballistic transmission in arbitrarily twisted bilayer $\textrm{MoS}_2$.
\end{abstract}

%\pacs{Valid PACS appear here}% PACS, the Physics and Astronomy
                             % Classification Scheme.
%Use showkeys class option if keyword
                              %display desired
\maketitle

\section{Introduction }
\begin{figure}%[h]
  \centering
    \includegraphics[width=0.45\textwidth]{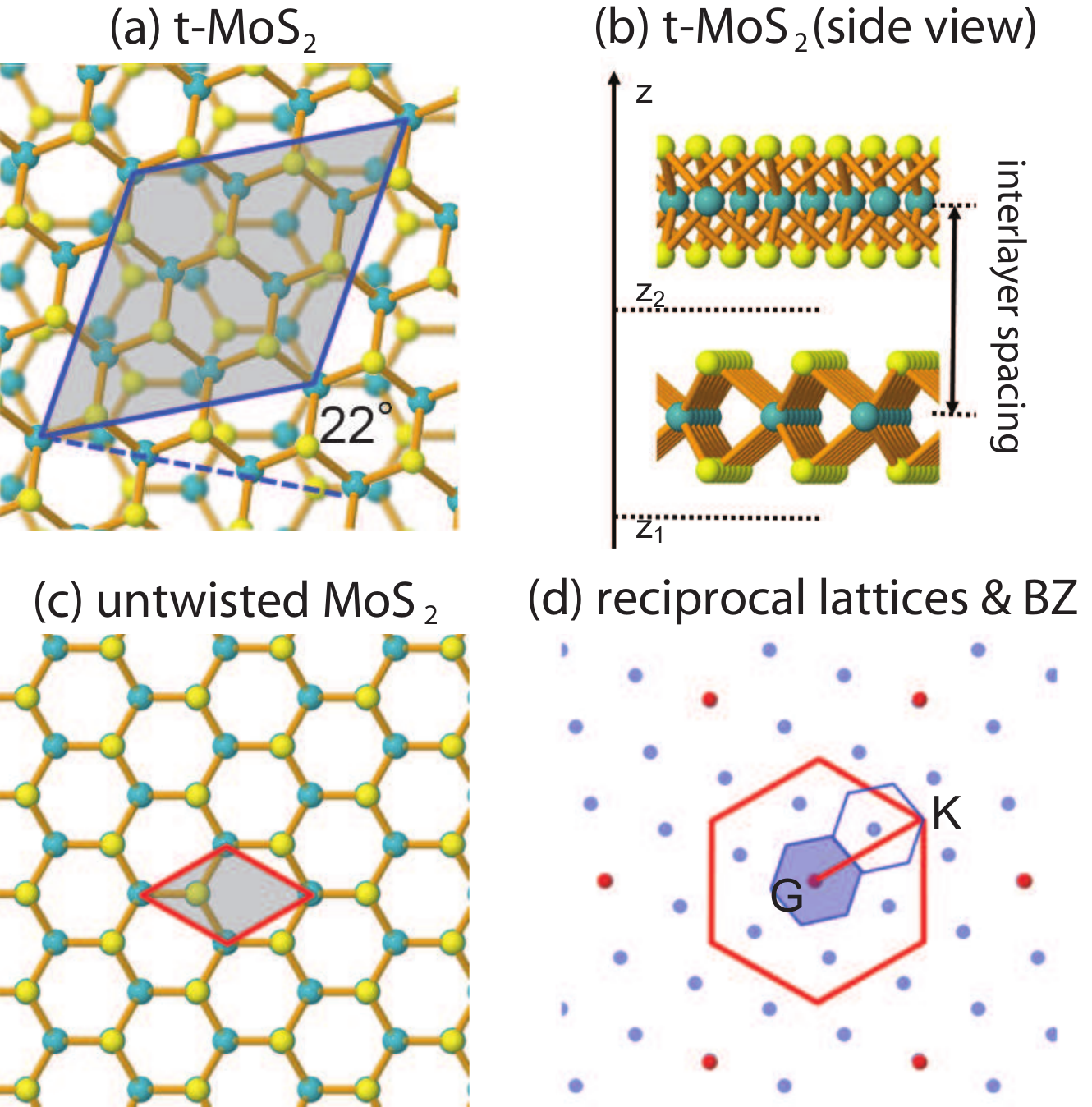}
    \caption{(a),(b) top view and sideview  of the supercell of a $\textrm{t-MoS}_2$ with a twist angle of $22^\circ$. (c) Primitive unit cell of un-twisted bilayer $\textrm{MoS}_2$. (d) Corresponding reciprocal lattices of the the twisted and untwisted structures. The unit cells are denoted by the shaded parallelograms.  The $\textrm{t-MoS}_2$ unit cell contains 42 atoms while the un-twisted bilayer $\textrm{MoS}_2$ contains 6 atoms. In (d), the solid blue and red hexagons correspond to the Brillouin zones (BZ) of twisted and untwisted bilayer $\textrm{MoS}_2$ respectively. Blue dots correspond to the reciprocal lattice vectors $\mathbf{G}_S$ of the supercell, while red dots correspond to the reciprocal lattice vectors $\mathbf{G}_P$ of the primitive unit cell. 
}    \label{fig:Unitcell_and_BZ}
\end{figure}

Two dimensional (2D) materials such as graphene and transition metal dichalcogenides (TMDs)
constitute exciting candidates for a variety of electronic and optoelectronic device applications \cite{Radisav_Single_layer_MoS2FET,Das_MultilayerMoS2_with_Scandium}. 
In particular, there is growing interest
in stacked 2D materials that often occur naturally during the growth process, and
also provide opportunities for added functionalities due to their varying thickness, crystal orientation and composition.
It is critical to understand the electronic properties of stacked 2D materials such as twisted multilayer TMDs and TMD heterostructures. However, the complexity arises because their electronic properties 
are highly sensitive to morphology and inter-layer interactions \cite{VanderWaals_heterostructure,twisted_bilayer_Arend,twisted_bilayer_Wang_theory,twisted_bilayer_Yeh}.
\\\\
The translational symmetry of a twisted multilayer TMD
is compromised because of its twist angle, requiring a
supercell that is considerably larger than the primitive unit cell (e.g. Fig.\ref{fig:Unitcell_and_BZ} (a)), and a corresponding convoluted spaghetti-like band structure due to the aggressive folding of its Brillouin zone (BZ) \cite{Boykin_BZ_unfolding}. 
In principle, atomistic first principles as well as empirical methods\cite{Hesam_TMD_FET,YaohuaTan_abinitioMapping} can  be used to model such twisted bilayer systems. However, as large unit cells of twisted systems increase the computational load dramatically, atomistic simulations are limited to specific twist angles with tractable BZ sizes instead of random orientations. A simpler model that nonetheless captures the essential physics of the bandedge and effective mass dependencies on orientation would be highly desirable. 
\\\\
In this work, the band structures of twisted bilayer $\textrm{MoS}_2$'s ($\textrm{t-MoS}_2$) are obtained using first principles calculations. In order to extract meaningful parameters such as bandedge splittings relevant to inter-layer interactions, the technique of band unfolding \cite{Boykin_BZ_unfolding,Boykin_BZ_unfolding_SiGe,Boykin_BZ_unfolding_alloy,
Popescu_unfolding,Paulo_unfolding} is applied to the twisted bilayer TMDs. Interlayer interactions are extracted from the unfolded band structures, and then phenomenological multi-valley effective mass models are constructed to model the indirect valence bands at the $\Gamma$ point and the indirect conduction bands along $\Gamma-K$ directions. 

\section{Method}\label{sec:method}
In this work, calculations based on Density
Functional Theory (DFT) are performed using the Projector-Augmented Wave
(PAW) technique implemented in the Vienna Ab-initio
Simulation Package (VASP)\cite{VASP_Kresse}. The
PBE functional\cite{PBE} is used to model the electron exchange-correlation.
The stacked TMD monolayers are weakly coupled through a Van der Waals force that is modeled using 
the VdW functional optB88 functional \cite{optB88_VdW,Dion_VdW}. The opt88 functional has been proved to be reliable for binding energies and geometries of Van der Waals structures such as graphite and h-BN contacted with metals \cite{Li_Graphene,BN_metal}. In all our calculations, a cut off energy of 400 eV  is used. A $2 \times 2 \times 1$ $\Gamma$-centered Monkhorst Pack kspace grid is used to describe the large $\textrm{t-MoS}_2$ supercells. For a $\textrm{t-MoS}_2$ structure with a smaller unitcell
(with twist angles of $0^\circ$ and $60^\circ$), a denser  $12 \times 12 \times 1$ k-space grid is used.
\\\\
To extract information from the massive number of bands of a $\textrm{t-MoS}_2$ system, we employ the technique of band unfolding\cite{Popescu_unfolding,Boykin_BZ_unfolding}. This technique allows us to unfold the bands in the Brillouin zone of the large supercell back into the Brillouin zone of the primitive unit cell of untwisted MoS$_2$. 
Fig. \ref{fig:Unitcell_and_BZ} shows the unit cell of a twisted and un-twisted bilayer $\textrm{MoS}_2$ and their corresponding Brillouin zones. The reciprocal vectors of the supercell and the primitive unit cell are denoted by $\mathbf{G}_S$ and $\mathbf{G}_P$ respectively, with the $\mathbf{G}_P$'s forming a subset of $\{ \mathbf{G}_S \}$.
For a twisted bilayer system, the primitive unit cell of either the upper or the lower layer can be used to unfold the band structures. In the band unfolding process, eigen states of the system in the supercell are decomposed into linear combination of Fourier components of the primitive unit cell 
\begin{equation}\label{eq:wf_unfold}
|\Psi_{n,\mathbf{k}_S}\rangle = \sum_{\mathbf{k}_P}a_{\mathbf{k}_P}|\Psi_{n,\mathbf{k}_P}\rangle.
\end{equation}
Here the $\mathbf{k}_S$ and $\mathbf{k}_P$ correspond respectively to wave vectors in the Brillouin zones of the supercell and the primitive unit cell. 
Each $\mathbf{k}_S$ can be unfolded to a few $\mathbf{k}_P $ satisfying $\mathbf{k}_P = \mathbf{k}_S + \mathbf{G}_S$. Taking the systems and their corresponding Brillouin zones shown in Fig.\ref{fig:Unitcell_and_BZ} as an example, the
unfolded $\mathbf{G}_S$ correspond to the blue dots enclosed by the Brillouin zone of the supercell. The resulting unfolded structures can now be readily compared to the original un-twisted system as both band structures lie in the Brillouin zone of the same primitive unit cell.
\\\\
For a bilayer system, the unfolded band structure contains bands contributed by both layers. 
To further separate the bands of one monolayer from another, the monolayer projector   can be used.
\begin{equation}\label{eq:wf_unfold_proj}
\hat{P}_{L}|\Psi_{n,\mathbf{k}_S}\rangle = \sum_{\mathbf{k}_P}a_{\mathbf{k}_P}\hat{P}_{L}|\Psi_{n,\mathbf{k}_P}\rangle,
\end{equation}
where the $\hat{P}_{L}$ is the monolayer projector defined as
\begin{equation}\label{eq:z_projector}
\hat{P}_{L} = |h(z) \rangle \langle h(z)|,\quad
h\left(z\right) =  \begin{cases}
\frac{1}{\sqrt{z_2-z_1}}, & z_1 \leq z \leq z_2 \\
0  & \textrm{otherwise}
\end{cases}.
\end{equation}
where $z_1$ and $z_2$ define the region of one of its layers. For instance, the $z_1$ and $z_2$ in Fig. \ref{fig:Unitcell_and_BZ} (b) define the region of the lower layer in the $\textrm{t-MoS}_2$. Half of the interlayer spacing is included in this definition.

\section{Results}\label{sec:results}
\begin{figure}[h]
  \centering
    \includegraphics[width=0.4\textwidth]{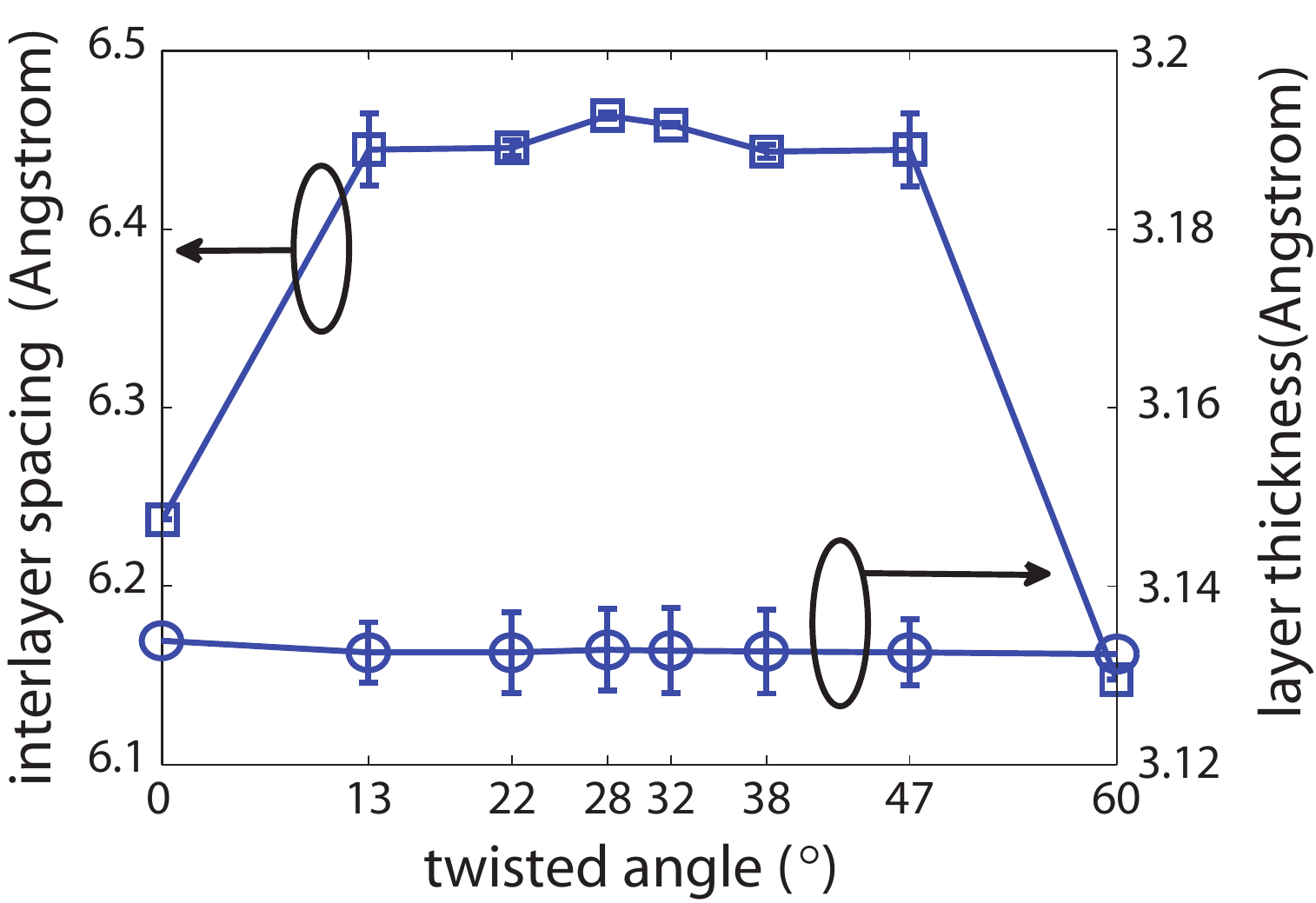}%
    \caption{Twist angle dependence of the thickness and layer spacing (distance between $\textrm{Mo}$ planes) of $\textrm{t-MoS}_2$. The average thickness of $\textrm{MoS}_2$ monolayers changes only slightly, while the interlayer spacing of the $\textrm{t-MoS}_2$ varies more prominently. We reach a maximum spacing for a twist angle of about $30^\circ$.  }    \label{fig:thickness_spacing}
\end{figure}

In this work, we consider $\textrm{t-MoS}_2$ with special twist angles ($0^\circ$, $13^\circ$, $22^\circ$,   $28^\circ$, $32^\circ$, $38^\circ$, $47^\circ$ and $60^\circ$). Of all these cases, the $\textrm{t-MoS}_2$ with the twist angles of $13^\circ$ and $47^\circ$ have the largest unit cell with 114 atoms (compared with just 6 atoms in a un-twisted bilayer $\textrm{MoS}_2$ system). For all the twisted structures, the lattice constant is chosen as $a=3.18\AA$ according to previous work by ref. \onlinecite{MoS2_GW_Vasp}.
Fig.\ref{fig:thickness_spacing} shows all the inter-layer distances and monolayer thicknesses plotted against the twist angle of $\textrm{t-MoS}_2$. We see that the average thickness of each layer in $\textrm{t-MoS}_2$ is weakly dependent on the twist angle. A weak inhomogenous strain is introduced in response to the broken translation symmetry in $\textrm{t-MoS}_2$. The variances of the layer thickness in all the considered systems are less than 0.01$\AA$, corresponding to a maximum diagonal strain component of $\varepsilon_{zz} = 0.3\%$. In contrast the inter-layer spacings, defined as the distance between the Mo-planes as it is shown in Fig.\ref{fig:Unitcell_and_BZ} (b), change more substantially with respect to the twist angle. Compared with un-twisted $\textrm{MoS}_2$, the inter-layer spacing increases by 0.2$\AA$ between twist angles from $13^\circ$ to $47^\circ$. 

\begin{figure*}%[h]
  \centering
    \includegraphics[width=\textwidth]{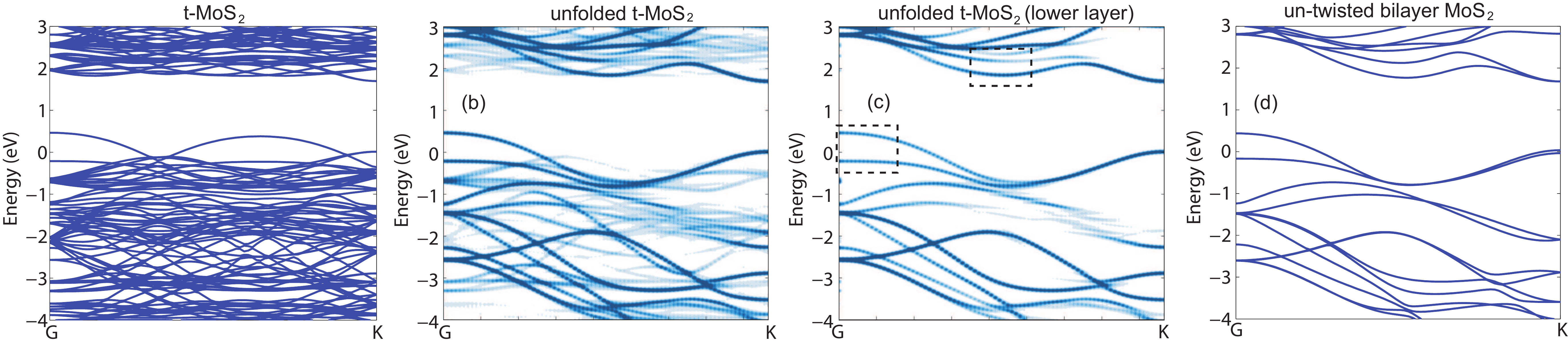}%
    \caption{(a) Folded and (b) unfolded band structures of  $\textrm{t-MoS}_2$. The weights $|a_{\mathbf{k}_P}|^2$ in eq.(\ref{eq:wf_unfold}) are represented by the intensity of the blue coloration. (c) Unfolded band structure contributed by the lower layer obtained by projecting ($|a_{\mathbf{k}_P}|^2 \langle \Psi_{n,\mathbf{k}_P}|\hat{P}_{L}|\Psi_{n,\mathbf{k}_P}\rangle$ in Eq.(\ref{eq:wf_unfold_proj}) is represented by the blue coloration). (d) Bandstructure of the original un-twisted bilayer $\textrm{MoS}_2$. In (a), the indirect conduction valleys are embedded in the huge number of sphagetti like bands due to Brillouin zone folding. With band unfolding, the interfering bands are filtered out and both direct and indirect valleys can be seen clearly in (b) and (c). In (c), the unfolded band structures are further projected on to the lower layer of the $\textrm{t-MoS}_2$. From (c), the indirect valleys marked by rectangles can be investigated in detail. 
Compared with (d), we readily see that some of the original untwisted bands are broken and broadened by interlayer interactions in the twisted bilayer system. }    \label{fig:band_structures}
\end{figure*}
\begin{figure}%[h]
  \centering\hfill
    \includegraphics[width=0.49\textwidth]{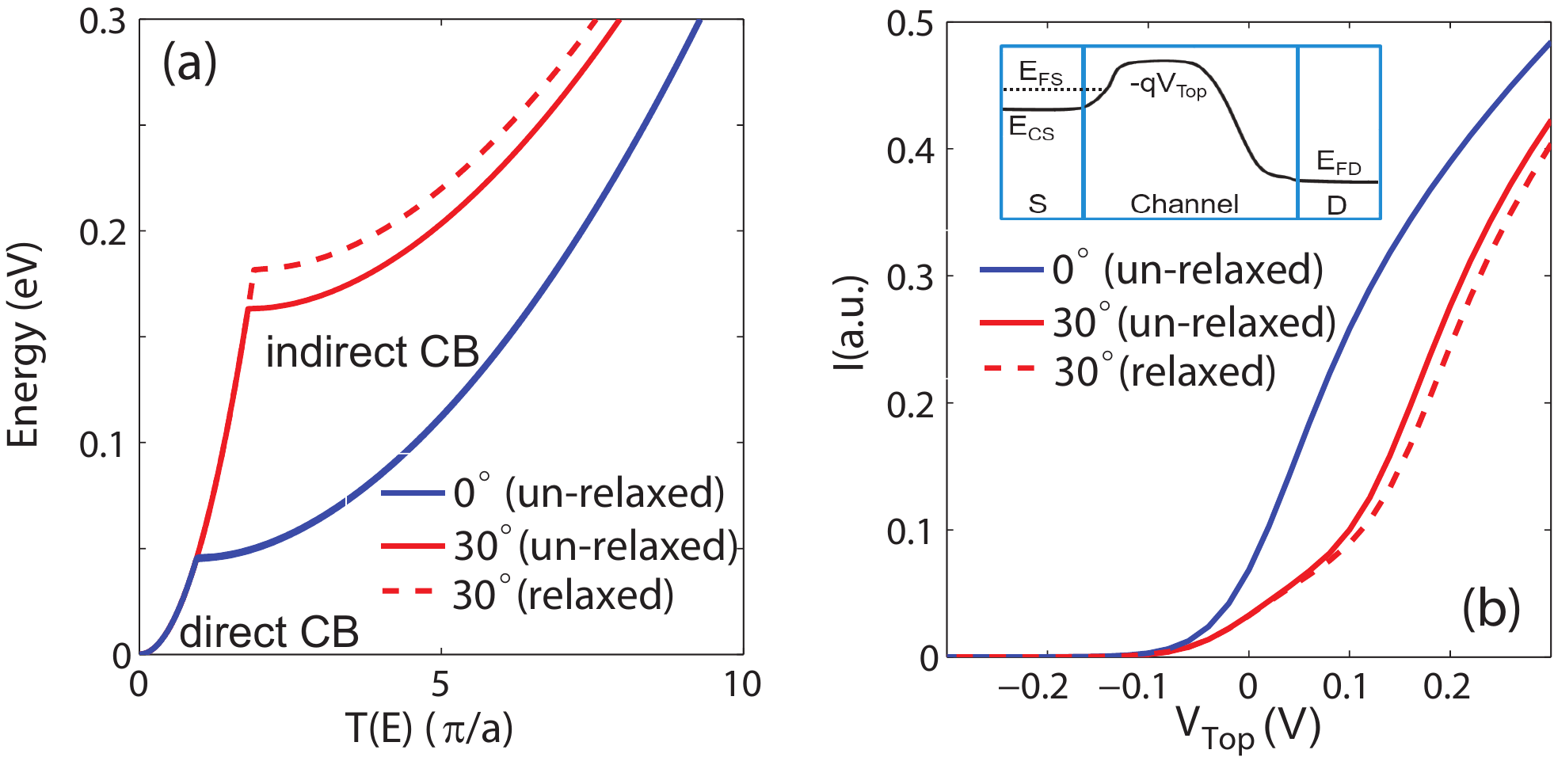}
    \caption{  (a) Integrated transmission of un-twisted and $\textrm{t-MoS}_2$. (b) Ballistic current in un-twisted and $\textrm{t-MoS}_2$ transistors. Compared with un-twisted  bilayer $\textrm{MoS}_2$, a 50\% reduction in on current is expected for the twisted bilayer  $\textrm{MoS}_2$ channel. Here the $E_{FS}-E_{FD} = 0.05eV$, with $0$ lying at the top of the barrier. 
     }    \label{fig:current}
\end{figure}

In order to understand the impact of the twist angle and the resulting variation in inter-layer distance on the band structures, we look closer at the band structures of two kinds of $\textrm{t-MoS}_2$ structures: 1. $\textrm{t-MoS}_2$  with a fixed inter-layer distance (no geometry relaxation is applied in this case); and 2. $\textrm{t-MoS}_2$ after geometry relaxation. For un-relaxed structures in case 1,  the inter-layer distance of all the $\textrm{t-MoS}_2$s is set equal to that of a relaxed, un-twisted $\textrm{MoS}_2$. For these $\textrm{t-MoS}_2$'s with a fixed inter-layer distance, the DFT band structures of $\textrm{t-MoS}_2$ with structures in Fig.\ref{fig:Unitcell_and_BZ} are shown in Fig. \ref{fig:band_structures}.
Compared with the band structure of the un-twisted bilayer $\textrm{MoS}_2$ in \ref{fig:band_structures}.(d), the band structure of a $\textrm{t-MoS}_2$ in Fig. \ref{fig:band_structures} (a) has little resemblance due to Brillouin zone folding. 
With band unfolding however, the interfering bands are filtered out. The unfolded band structure in Fig. \ref{fig:band_structures} (b) is comparable with the band structure of un-twisted bilayer $\textrm{MoS}_2$ in Fig.\ref{fig:band_structures}.(c). 
The probability amplitude of each Fourier component $|\Psi_{n,\mathbf{k}_P}\rangle$ given by $|a_{\mathbf{k}_P}|^2$ in equation (\ref{eq:wf_unfold}) is represented by the color intensity in \ref{fig:band_structures}.(b).
In Fig. \ref{fig:band_structures}.(c), the bands in Fig. \ref{fig:band_structures}.(b) are further projected to the lower layer by applying the monolayer projector $\hat{P}_L$ given by equation (\ref{eq:z_projector}). The probability amplitudes $|a_{\mathbf{k}_P}|^2 \langle \Psi_{n,\mathbf{k}_P}|\hat{P}_L|\Psi_{n,\mathbf{k}_P}\rangle$ are shown in  Fig.\ref{fig:band_structures}.(c).
Compared with the band structure in \ref{fig:band_structures}.(d), we clearly see how broken bands and broadened bands appear in Fig. \ref{fig:band_structures}.(b) due to the inter-layer interactions in the twisted bilayer system. 
\begin{figure}%[h]
  \centering\hfill  
    \includegraphics[width=0.49\textwidth]{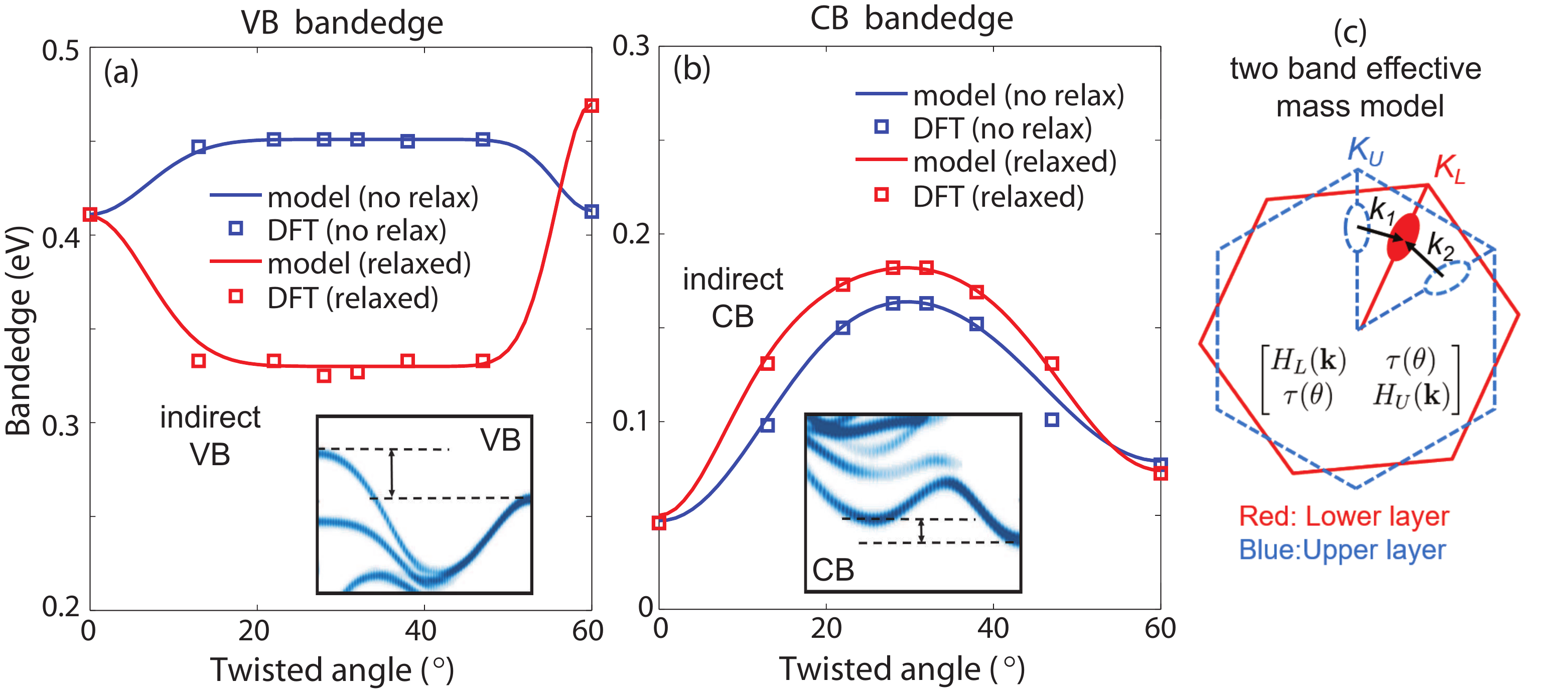}
    \caption{(a) Twist angle dependent indirect VB  and (b) indirect CB bandedges - comparing the unfolded DFT band-edgees with our fitted and parametrized two band model (Eqs.~\ref{eq:model},~\ref{eq:model2} and table I). 
    (c) Summary of two band effective mass model for indirect conduction bands. The simple model shows excellent agreement with DFT results for all the angles studied.}
    \label{fig:bandedge}
\end{figure}
\\\\
For other $\textrm{t-MoS}_2$ systems, the unfolding process is applied analogously, and figures similar to Fig. \ref{fig:band_structures}.(c) are obtained in each case. The bandedges and band splittings of important valleys are also quantitatively evaluated from the unfolded bands of relaxed and un-relaxed $\textrm{t-MoS}_2$, as marked by rectangles in Fig.\ref{fig:band_structures}.(c).
For the relaxed $\textrm{t-MoS}_2$ systems, the small thickness variations within the supercell of the twisted bilayer systems are seen to induce only a negligible variation to the band gaps and effective masses. Such a weak variation is expected because the average strain $\varepsilon_{zz}$ of each layer is negligible in the $\textrm{t-MoS}_2$.  For the direct conduction and valence bands at the $K$ point, the changes in bandedges due to twist angle are also negligible. However, the indirect conduction and valence valleys are seen to have a stronger dependence on interlayer interactions. 
The uppermost valence bands (VB) of relaxed $\textrm{t-MoS}_2$'s are about 0.12eV lower than that of un-relaxed $\textrm{t-MoS}_2$'s for twist angles running between
$15^\circ$ to $45^\circ$, while for a $60^\circ$ twist angle, it rises above the un-relaxed value. 
The variation of VB at the $\Gamma$ point reaches about 0.1eV, which quanlitatively agrees with measurements by Ref \onlinecite{twisted_bilayer_Yeh}. These deviations suggest that both the twist angle and the change in thickness have considerable impact on the bandedge of the indirect valence band at the $\Gamma$ point. 
The indirect conduction bands (CBs) in relaxed $\textrm{t-MoS}_2$ are slightly higher than that of the un-relaxed $\textrm{t-MoS}_2$. The maximum discrepancy is about 0.04 eV for a  twist angle of about $30^\circ$, 
suggesting that the twist angle also influences the bandedges of indirect CBs. 
For both relaxed and un-relaxed structures, the indirect conduction valleys reach the highest energy (0.15eV above lowest conduction band) at a twist angle of $30^\circ$. For all these cases, the original un-twisted bilayer $\textrm{MoS}_2$ has the lowest indirect conduction valleys.
\begin{table}\label{tab:parameter}
  \centering
    \begin{tabular}{ccccccc}
     \hline
      % after \\: \hline or \cline{col1-col2} \cline{col3-col4} ...
     Valley & $a (eV)$ & $b(eV)$ & $c(eV)$ & $\sigma(\pi/3)$ & $\rho(\pi/3)$ \\
     \hline
    CB(relaxed)  & 0.131 &  0.067 &  0.041 & 0.172 & 0.215 \\
    CB(un-relaxed)  & 0.139 & 0.060 & 0.033 & 0.299 & 0.485  \\
    VB(relaxed)  & 0.199 & 0.082 & 0.140 & 0.122 & 0.113 \\
    VB(un-relaxed)  & 0.321 & -0.040 & -0.039 & 0.153 & 0.094 \\
    \hline
    \end{tabular}    
    \caption{ Parameters of the two band effective mass model. Effective masses of indirect valley are $m_l=0.601$ and $m_t=1.304$   }
\end{table}
\\\\
Using the bandedges extracted from the unfolded band structure, we can develop effective mass models to describe the angle dependent indirect CBs and VBs. For a twisted bilayer system with a twist angle $\theta$, we introduce a two band effective mass model to couple the indirect valleys of different layers. The two band effective mass model has a general expression given by
\begin{equation}\label{eq:model}
H = \left[ 
\begin{array}{cc}
H_L(\mathbf{k})& \tau\left( \theta \right) \\
 \tau\left( \theta \right)  & H_U\left(\mathbf{k}\right)
\end{array}
\right].
\end{equation}
where $H_U\left(\mathbf{k}\right)$ and $H_L\left(\mathbf{k}\right)$
correspond to the valleys of upper and lower layers in a $\textrm{t-MoS}_2$ system.
The twist angle dependent $\tau(\theta)$ is parameterized with an analytical expression 
\begin{equation}\label{eq:model2}
\tau(\theta) =  a + b\exp{\left(-\frac{\theta^2}{\sigma^2}\right)} + c\exp{\left[ -\frac{(\theta - {\pi}/{3} )^2}{\rho^2}\right]} , 
\end{equation}
where $a,b,c,\sigma$ and $\rho$ are fitting parameters. This expression of $\tau(\theta)$ is valid for $0 \leq\theta \leq \pi/3$, with the relations $\tau(\theta+2n\pi/3) =\tau(\theta)$ and $\tau(2\pi/3-\theta) =\tau(\theta)$ beyond that limit.
For valence valleys at the $\Gamma$ point, we have 
$H_L\left(\mathbf{k}\right) = H_U\left(\mathbf{k}\right) = {\hbar^2k^2}/{2m^*}$. 
For indirect CBs, the twist angle corresponds to a change in the bottom of the conduction valleys in the Brillouin zone, as shown in Fig. \ref{fig:bandedge} (c).  
For uncoupled CBs along the k-direction with an angle $\Omega$,
the un-coupled CB valley is given by a generalized effective mass model
\begin{eqnarray}
E(\mathbf{k},\Omega) & = &{\hbar^2} \left( k^2_x/2m_{xx} +k_xk_y/m_{xy} + k^2_y/2m_{yy} \right)\\
\nonumber 1/m_{xx}& = &\cos^2{\Omega}/m_l + \sin^2{\Omega}/m_t \\ 
\nonumber 1/m_{yy}& = &\sin^2{\Omega}/m_l + \cos^2{\Omega}/m_t \\
\nonumber 1/m_{xy}& = &\left(1/m_t -1/m_l\right) \sin{\Omega}\cos{\Omega}
\end{eqnarray}
\\\\
Consider one of the valleys from the lower layer, whose Hamiltonian $H_L = E(\mathbf{k},\Omega)$.
Two of the valleys from the upper layer can interact with this $H_L$. 
Based on our simulations we model the upper valleys by
$H_U = E_U+E(\mathbf{k},\Omega)$ and $E_U = E(\mathbf{k}_1,\Omega_1) E(\mathbf{k}_2,\Omega_2) / ( E(\mathbf{k}_1,\Omega_1)+ E(\mathbf{k}_2,\Omega_2))$. 
The parameters of indirect valence and conduction valleys are listed in table. \ref{tab:parameter}.
From Fig.\ref{fig:bandedge}, it can be seen that the two band effective mass models introduced in this work reproduce the DFT bandedges accurately.
\\\\
%\end{itemize}
Figure \ref{fig:current}.(a) shows the integrated electron transmission across $\textrm{t-MoS}_2$ with different twist angles. We see that the indirect conduction bands have higher energy than the direct conduction bands, as shown in Fig.\ref{fig:bandedge} (b). However the indirect conduction bands contribute a larger integrated transmission because they have larger effective masses and because there are six valleys. In Fig.\ref{fig:current} (b), the ballistic currents in n-type twisted and untwisted bilayer $\textrm{MoS}_2$ are calculated using a top of barrier model\cite{TopofBarrier}. We see a reduction in ON current of up to 50\% for the twisted bilayer  $\textrm{MoS}_2$ channel, suggesting that the indirect conduction bands bear significant contribution to the device performance.

%\pagebreak
\section{Conclusion}\label{sec:Conclusion}
In summary, a Brillouin zone unfolding technique is employed to explore the electronic band structures and ballistic current flow in twisted bilayer MoS2. A simple, twist angle-dependent
two band effective mass model is developed to model the bandedges of the indirect valleys. Such an angle-dependent parametrization can in principle allow a configurational average over the twist angle distribution to properly quantify the role of configurational disorder in 2D layered materials. 

\begin{acknowledgments}
This project was supported by the Nano Research Initiative (NRI) through the Institute for Nanoelectronics Discovery and Exploration (INDEX) center.
\end{acknowledgments}

%\appendix
%
%\section{Expression of $\mathcal{M}^{(l)}_{\alpha,\gamma}\left(\mathbf{\hat{d}}\right)$}\label{app:D_matrix}

%\bibliography{apssamp}% Produces the bibliography via BibTeX.

\begin{thebibliography}{20}
\expandafter\ifx\csname natexlab\endcsname\relax\def\natexlab#1{#1}\fi
\expandafter\ifx\csname bibnamefont\endcsname\relax
  \def\bibnamefont#1{#1}\fi
\expandafter\ifx\csname bibfnamefont\endcsname\relax
  \def\bibfnamefont#1{#1}\fi
\expandafter\ifx\csname citenamefont\endcsname\relax
  \def\citenamefont#1{#1}\fi
\expandafter\ifx\csname url\endcsname\relax
  \def\url#1{\texttt{#1}}\fi
\expandafter\ifx\csname urlprefix\endcsname\relax\def\urlprefix{URL }\fi
\providecommand{\bibinfo}[2]{#2}
\providecommand{\eprint}[2][]{\url{#2}}

\bibitem[{\citenamefont{Radisavljevic et~al.}(2011)\citenamefont{Radisavljevic,
  Radenovic, Brivio, Giacometti, and Kis}}]{Radisav_Single_layer_MoS2FET}
\bibinfo{author}{\bibfnamefont{B.}~\bibnamefont{Radisavljevic}},
  \bibinfo{author}{\bibfnamefont{A.}~\bibnamefont{Radenovic}},
  \bibinfo{author}{\bibfnamefont{J.}~\bibnamefont{Brivio}},
  \bibinfo{author}{\bibfnamefont{V.}~\bibnamefont{Giacometti}},
  \bibnamefont{and} \bibinfo{author}{\bibfnamefont{A.}~\bibnamefont{Kis}},
  \bibinfo{journal}{Nature Nanotechnology} \textbf{\bibinfo{volume}{6}},
  \bibinfo{pages}{147} (\bibinfo{year}{2011}).

\bibitem[{\citenamefont{Das et~al.}(2013)\citenamefont{Das, Chen, Penumatcha,
  and Appenzeller}}]{Das_MultilayerMoS2_with_Scandium}
\bibinfo{author}{\bibfnamefont{S.}~\bibnamefont{Das}},
  \bibinfo{author}{\bibfnamefont{H.-Y.} \bibnamefont{Chen}},
  \bibinfo{author}{\bibfnamefont{A.~V.} \bibnamefont{Penumatcha}},
  \bibnamefont{and}
  \bibinfo{author}{\bibfnamefont{J.}~\bibnamefont{Appenzeller}},
  \bibinfo{journal}{Nano Letters} \textbf{\bibinfo{volume}{13}},
  \bibinfo{pages}{100} (\bibinfo{year}{2013}), \bibinfo{note}{pMID: 23240655},
  \eprint{http://dx.doi.org/10.1021/nl303583v},
  \urlprefix\url{http://dx.doi.org/10.1021/nl303583v}.

\bibitem[{\citenamefont{Fang et~al.}(2014)\citenamefont{Fang, Battaglia,
  Carraro, Nemsak, Ozdol, Kang, Bechtel, Desai, Kronast, Unal
  et~al.}}]{VanderWaals_heterostructure}
\bibinfo{author}{\bibfnamefont{H.}~\bibnamefont{Fang}},
  \bibinfo{author}{\bibfnamefont{C.}~\bibnamefont{Battaglia}},
  \bibinfo{author}{\bibfnamefont{C.}~\bibnamefont{Carraro}},
  \bibinfo{author}{\bibfnamefont{S.}~\bibnamefont{Nemsak}},
  \bibinfo{author}{\bibfnamefont{B.}~\bibnamefont{Ozdol}},
  \bibinfo{author}{\bibfnamefont{J.~S.} \bibnamefont{Kang}},
  \bibinfo{author}{\bibfnamefont{H.~A.} \bibnamefont{Bechtel}},
  \bibinfo{author}{\bibfnamefont{S.~B.} \bibnamefont{Desai}},
  \bibinfo{author}{\bibfnamefont{F.}~\bibnamefont{Kronast}},
  \bibinfo{author}{\bibfnamefont{A.~A.} \bibnamefont{Unal}},
  \bibnamefont{et~al.}, \bibinfo{journal}{Proceedings of the National Academy
  of Sciences} \textbf{\bibinfo{volume}{111}}, \bibinfo{pages}{6198}
  (\bibinfo{year}{2014}),
  \eprint{http://www.pnas.org/content/111/17/6198.full.pdf},
  \urlprefix\url{http://www.pnas.org/content/111/17/6198.abstract}.

\bibitem[{\citenamefont{van~der Zande et~al.}(2014)\citenamefont{van~der Zande,
  Kunstmann, Chernikov, Chenet, You, Zhang, Huang, Berkelbach, Wang, Zhang
  et~al.}}]{twisted_bilayer_Arend}
\bibinfo{author}{\bibfnamefont{A.~M.} \bibnamefont{van~der Zande}},
  \bibinfo{author}{\bibfnamefont{J.}~\bibnamefont{Kunstmann}},
  \bibinfo{author}{\bibfnamefont{A.}~\bibnamefont{Chernikov}},
  \bibinfo{author}{\bibfnamefont{D.~A.} \bibnamefont{Chenet}},
  \bibinfo{author}{\bibfnamefont{Y.}~\bibnamefont{You}},
  \bibinfo{author}{\bibfnamefont{X.}~\bibnamefont{Zhang}},
  \bibinfo{author}{\bibfnamefont{P.~Y.} \bibnamefont{Huang}},
  \bibinfo{author}{\bibfnamefont{T.~C.} \bibnamefont{Berkelbach}},
  \bibinfo{author}{\bibfnamefont{L.}~\bibnamefont{Wang}},
  \bibinfo{author}{\bibfnamefont{F.}~\bibnamefont{Zhang}},
  \bibnamefont{et~al.}, \bibinfo{journal}{Nano Letters}
  \textbf{\bibinfo{volume}{14}}, \bibinfo{pages}{3869} (\bibinfo{year}{2014}),
  \bibinfo{note}{pMID: 24933687},
  \urlprefix\url{http://dx.doi.org/10.1021/nl501077m}.

\bibitem[{\citenamefont{Wang et~al.}(2015)\citenamefont{Wang, Chen, and
  Wang}}]{twisted_bilayer_Wang_theory}
\bibinfo{author}{\bibfnamefont{Z.}~\bibnamefont{Wang}},
  \bibinfo{author}{\bibfnamefont{Q.}~\bibnamefont{Chen}}, \bibnamefont{and}
  \bibinfo{author}{\bibfnamefont{J.}~\bibnamefont{Wang}}, \bibinfo{journal}{The
  Journal of Physical Chemistry C} \textbf{\bibinfo{volume}{119}},
  \bibinfo{pages}{4752} (\bibinfo{year}{2015}),
  \urlprefix\url{http://dx.doi.org/10.1021/jp507751p}.

\bibitem[{\citenamefont{Yeh et~al.}(2016)\citenamefont{Yeh, Jin, Zaki,
  Kunstmann, Chenet, Arefe, Sadowski, Dadap, Sutter, Hone
  et~al.}}]{twisted_bilayer_Yeh}
\bibinfo{author}{\bibfnamefont{P.-C.} \bibnamefont{Yeh}},
  \bibinfo{author}{\bibfnamefont{W.}~\bibnamefont{Jin}},
  \bibinfo{author}{\bibfnamefont{N.}~\bibnamefont{Zaki}},
  \bibinfo{author}{\bibfnamefont{J.}~\bibnamefont{Kunstmann}},
  \bibinfo{author}{\bibfnamefont{D.}~\bibnamefont{Chenet}},
  \bibinfo{author}{\bibfnamefont{G.}~\bibnamefont{Arefe}},
  \bibinfo{author}{\bibfnamefont{J.~T.} \bibnamefont{Sadowski}},
  \bibinfo{author}{\bibfnamefont{J.~I.} \bibnamefont{Dadap}},
  \bibinfo{author}{\bibfnamefont{P.}~\bibnamefont{Sutter}},
  \bibinfo{author}{\bibfnamefont{J.}~\bibnamefont{Hone}}, \bibnamefont{et~al.},
  \bibinfo{journal}{Nano Letters} \textbf{\bibinfo{volume}{16}},
  \bibinfo{pages}{953} (\bibinfo{year}{2016}), \bibinfo{note}{pMID: 26760447},
  \urlprefix\url{http://dx.doi.org/10.1021/acs.nanolett.5b03883}.

\bibitem[{\citenamefont{Boykin and Klimeck}(2005)}]{Boykin_BZ_unfolding}
\bibinfo{author}{\bibfnamefont{T.~B.} \bibnamefont{Boykin}} \bibnamefont{and}
  \bibinfo{author}{\bibfnamefont{G.}~\bibnamefont{Klimeck}},
  \bibinfo{journal}{Phys. Rev. B} \textbf{\bibinfo{volume}{71}},
  \bibinfo{pages}{115215} (\bibinfo{year}{2005}),
  \urlprefix\url{http://link.aps.org/doi/10.1103/PhysRevB.71.115215}.

\bibitem[{\citenamefont{Ilatikhameneh et~al.}(2015)\citenamefont{Ilatikhameneh,
  Tan, Novakovic, Klimeck, Rahman, and Appenzeller}}]{Hesam_TMD_FET}
\bibinfo{author}{\bibfnamefont{H.}~\bibnamefont{Ilatikhameneh}},
  \bibinfo{author}{\bibfnamefont{Y.}~\bibnamefont{Tan}},
  \bibinfo{author}{\bibfnamefont{B.}~\bibnamefont{Novakovic}},
  \bibinfo{author}{\bibfnamefont{G.}~\bibnamefont{Klimeck}},
  \bibinfo{author}{\bibfnamefont{R.}~\bibnamefont{Rahman}}, \bibnamefont{and}
  \bibinfo{author}{\bibfnamefont{J.}~\bibnamefont{Appenzeller}},
  \bibinfo{journal}{IEEE Journal on Exploratory Solid-State Computational
  Devices and Circuits} \textbf{\bibinfo{volume}{1}}, \bibinfo{pages}{12}
  (\bibinfo{year}{2015}), ISSN \bibinfo{issn}{2329-9231}.

\bibitem[{\citenamefont{Tan et~al.}(2015)\citenamefont{Tan, Povolotskyi, Kubis,
  Boykin, and Klimeck}}]{YaohuaTan_abinitioMapping}
\bibinfo{author}{\bibfnamefont{Y.~P.} \bibnamefont{Tan}},
  \bibinfo{author}{\bibfnamefont{M.}~\bibnamefont{Povolotskyi}},
  \bibinfo{author}{\bibfnamefont{T.}~\bibnamefont{Kubis}},
  \bibinfo{author}{\bibfnamefont{T.~B.} \bibnamefont{Boykin}},
  \bibnamefont{and} \bibinfo{author}{\bibfnamefont{G.}~\bibnamefont{Klimeck}},
  \bibinfo{journal}{Phys. Rev. B} \textbf{\bibinfo{volume}{92}},
  \bibinfo{pages}{085301} (\bibinfo{year}{2015}),
  \urlprefix\url{http://link.aps.org/doi/10.1103/PhysRevB.92.085301}.

\bibitem[{\citenamefont{Boykin et~al.}(2007{\natexlab{a}})\citenamefont{Boykin,
  Kharche, and Klimeck}}]{Boykin_BZ_unfolding_SiGe}
\bibinfo{author}{\bibfnamefont{T.~B.} \bibnamefont{Boykin}},
  \bibinfo{author}{\bibfnamefont{N.}~\bibnamefont{Kharche}}, \bibnamefont{and}
  \bibinfo{author}{\bibfnamefont{G.}~\bibnamefont{Klimeck}},
  \bibinfo{journal}{Phys. Rev. B} \textbf{\bibinfo{volume}{76}},
  \bibinfo{pages}{035310} (\bibinfo{year}{2007}{\natexlab{a}}),
  \urlprefix\url{http://link.aps.org/doi/10.1103/PhysRevB.76.035310}.

\bibitem[{\citenamefont{Boykin et~al.}(2007{\natexlab{b}})\citenamefont{Boykin,
  Kharche, Klimeck, and Korkusinski}}]{Boykin_BZ_unfolding_alloy}
\bibinfo{author}{\bibfnamefont{T.~B.} \bibnamefont{Boykin}},
  \bibinfo{author}{\bibfnamefont{N.}~\bibnamefont{Kharche}},
  \bibinfo{author}{\bibfnamefont{G.}~\bibnamefont{Klimeck}}, \bibnamefont{and}
  \bibinfo{author}{\bibfnamefont{M.}~\bibnamefont{Korkusinski}},
  \bibinfo{journal}{Journal of Physics: Condensed Matter}
  \textbf{\bibinfo{volume}{19}}, \bibinfo{pages}{036203}
  (\bibinfo{year}{2007}{\natexlab{b}}),
  \urlprefix\url{http://stacks.iop.org/0953-8984/19/i=3/a=036203}.

\bibitem[{\citenamefont{Popescu and Zunger}(2012)}]{Popescu_unfolding}
\bibinfo{author}{\bibfnamefont{V.}~\bibnamefont{Popescu}} \bibnamefont{and}
  \bibinfo{author}{\bibfnamefont{A.}~\bibnamefont{Zunger}},
  \bibinfo{journal}{Phys. Rev. B} \textbf{\bibinfo{volume}{85}},
  \bibinfo{pages}{085201} (\bibinfo{year}{2012}),
  \urlprefix\url{http://link.aps.org/doi/10.1103/PhysRevB.85.085201}.

\bibitem[{\citenamefont{Medeiros et~al.}(2014)\citenamefont{Medeiros,
  Stafstr\"om, and Bj\"ork}}]{Paulo_unfolding}
\bibinfo{author}{\bibfnamefont{P.~V.~C.} \bibnamefont{Medeiros}},
  \bibinfo{author}{\bibfnamefont{S.}~\bibnamefont{Stafstr\"om}},
  \bibnamefont{and} \bibinfo{author}{\bibfnamefont{J.}~\bibnamefont{Bj\"ork}},
  \bibinfo{journal}{Phys. Rev. B} \textbf{\bibinfo{volume}{89}},
  \bibinfo{pages}{041407} (\bibinfo{year}{2014}),
  \urlprefix\url{http://link.aps.org/doi/10.1103/PhysRevB.89.041407}.

\bibitem[{\citenamefont{Kresse and Furthmuller}(1996)}]{VASP_Kresse}
\bibinfo{author}{\bibfnamefont{G.}~\bibnamefont{Kresse}} \bibnamefont{and}
  \bibinfo{author}{\bibfnamefont{J.}~\bibnamefont{Furthmuller}},
  \bibinfo{journal}{Computational Materials Science}
  \textbf{\bibinfo{volume}{6}}, \bibinfo{pages}{15 } (\bibinfo{year}{1996}),
  ISSN \bibinfo{issn}{0927-0256}.

\bibitem[{\citenamefont{Perdew et~al.}(1996)\citenamefont{Perdew, Burke, and
  Ernzerhof}}]{PBE}
\bibinfo{author}{\bibfnamefont{J.~P.} \bibnamefont{Perdew}},
  \bibinfo{author}{\bibfnamefont{K.}~\bibnamefont{Burke}}, \bibnamefont{and}
  \bibinfo{author}{\bibfnamefont{M.}~\bibnamefont{Ernzerhof}},
  \bibinfo{journal}{Phys. Rev. Lett.} \textbf{\bibinfo{volume}{77}},
  \bibinfo{pages}{3865} (\bibinfo{year}{1996}).

\bibitem[{\citenamefont{Klime\ifmmode~\check{s}\else \v{s}\fi{}
  et~al.}(2011)\citenamefont{Klime\ifmmode~\check{s}\else \v{s}\fi{}, Bowler,
  and Michaelides}}]{optB88_VdW}
\bibinfo{author}{\bibfnamefont{J.~c.~v.}
  \bibnamefont{Klime\ifmmode~\check{s}\else \v{s}\fi{}}},
  \bibinfo{author}{\bibfnamefont{D.~R.} \bibnamefont{Bowler}},
  \bibnamefont{and}
  \bibinfo{author}{\bibfnamefont{A.}~\bibnamefont{Michaelides}},
  \bibinfo{journal}{Phys. Rev. B} \textbf{\bibinfo{volume}{83}},
  \bibinfo{pages}{195131} (\bibinfo{year}{2011}),
  \urlprefix\url{http://link.aps.org/doi/10.1103/PhysRevB.83.195131}.

\bibitem[{\citenamefont{Dion et~al.}(2004)\citenamefont{Dion, Rydberg,
  Schr\"oder, Langreth, and Lundqvist}}]{Dion_VdW}
\bibinfo{author}{\bibfnamefont{M.}~\bibnamefont{Dion}},
  \bibinfo{author}{\bibfnamefont{H.}~\bibnamefont{Rydberg}},
  \bibinfo{author}{\bibfnamefont{E.}~\bibnamefont{Schr\"oder}},
  \bibinfo{author}{\bibfnamefont{D.~C.} \bibnamefont{Langreth}},
  \bibnamefont{and} \bibinfo{author}{\bibfnamefont{B.~I.}
  \bibnamefont{Lundqvist}}, \bibinfo{journal}{Phys. Rev. Lett.}
  \textbf{\bibinfo{volume}{92}}, \bibinfo{pages}{246401}
  (\bibinfo{year}{2004}),
  \urlprefix\url{http://link.aps.org/doi/10.1103/PhysRevLett.92.246401}.

\bibitem[{\citenamefont{Hazrati et~al.}(2014)\citenamefont{Hazrati, de~Wijs,
  and Brocks}}]{Li_Graphene}
\bibinfo{author}{\bibfnamefont{E.}~\bibnamefont{Hazrati}},
  \bibinfo{author}{\bibfnamefont{G.~A.} \bibnamefont{de~Wijs}},
  \bibnamefont{and} \bibinfo{author}{\bibfnamefont{G.}~\bibnamefont{Brocks}},
  \bibinfo{journal}{Phys. Rev. B} \textbf{\bibinfo{volume}{90}},
  \bibinfo{pages}{155448} (\bibinfo{year}{2014}),
  \urlprefix\url{http://link.aps.org/doi/10.1103/PhysRevB.90.155448}.

\bibitem[{\citenamefont{Bokdam et~al.}(2014)\citenamefont{Bokdam, Brocks,
  Katsnelson, and Kelly}}]{BN_metal}
\bibinfo{author}{\bibfnamefont{M.}~\bibnamefont{Bokdam}},
  \bibinfo{author}{\bibfnamefont{G.}~\bibnamefont{Brocks}},
  \bibinfo{author}{\bibfnamefont{M.~I.} \bibnamefont{Katsnelson}},
  \bibnamefont{and} \bibinfo{author}{\bibfnamefont{P.~J.} \bibnamefont{Kelly}},
  \bibinfo{journal}{Phys. Rev. B} \textbf{\bibinfo{volume}{90}},
  \bibinfo{pages}{085415} (\bibinfo{year}{2014}),
  \urlprefix\url{http://link.aps.org/doi/10.1103/PhysRevB.90.085415}.

\bibitem[{\citenamefont{Ramasubramaniam}(2012)}]{MoS2_GW_Vasp}
\bibinfo{author}{\bibfnamefont{A.}~\bibnamefont{Ramasubramaniam}},
  \bibinfo{journal}{Phys. Rev. B} \textbf{\bibinfo{volume}{86}},
  \bibinfo{pages}{115409} (\bibinfo{year}{2012}),
  \urlprefix\url{http://link.aps.org/doi/10.1103/PhysRevB.86.115409}.

\end{thebibliography}

\end{document}